\newcommand{\be}{\begin{eqnarray}}
\newcommand{\ee}{\end{eqnarray}}

\newcommand{\bea}{\begin{eqnarray}}
\newcommand{\eea}{\end{eqnarray}}
\newcommand{\beas}{\begin{eqnarray*}}
\newcommand{\eeas}{\end{eqnarray*}}
\documentclass[twocolumn,aps,showpacs]{revtex4}
\usepackage{graphics}
\usepackage{epsfig}
\usepackage{amsmath}
\usepackage{amssymb}
\usepackage{amsthm}
\begin{document}
\title{Dynamical quark recombination in ultrarelativistic heavy-ion
  collisions and the proton to pion ratio}  
\author{Alejandro Ayala$^\dagger$, Mauricio Mart\1nez$^\dagger$, 
Guy Pai\'c$^\dagger$ and G. Toledo S\'anchez$^*$}   
\affiliation{$^\dagger$Instituto de Ciencias Nucleares, Universidad
Nacional Aut\'onoma de M\'exico, Apartado Postal 70-543, M\'exico
Distrito Federal 04510, Mexico.\\
$^*$Instituto de F\1sica, Universidad
Nacional Aut\'onoma de M\'exico, Apartado Postal 20-364, M\'exico
Distrito Federal 01000, Mexico.}

\begin{abstract}

We study quark thermal recombination as a function of energy density during
the evolution of a heavy-ion collision in a numerical model that reproduces
aspects of QCD phenomenology. We show that starting with a set of free quarks
(or quarks and antiquarks) the probability to form colorless clusters of
three quarks differs from that to form colorless clusters of quark-antiquark
and that the former has a sharp jump at a critical energy density whereas the
latter transits smoothly from the low to the high energy density domains. We
interpret this as a quantitative difference in the production of baryons and
mesons with energy density. We use this approach to compute the proton and pion
spectra in a Bjorken scenario that incorporates the evolution of these
probabilities with energy density, and therefore with proper time. From the
spectra, we compute the proton to pion ratio and compare to data at the
highest RHIC energies. We show that for a standard choice of parameters, this
ratio reaches one, though the maximum is very sensitive to the initial
evolution proper time.

\end{abstract}


\maketitle

\date{\today}

\section{Introduction}

The features of the proton to pion ratio for Au + Au collisions at the highest
energy at RHIC~\cite{PHENIXBM} have been interpreted as the
emergence of thermal recombination as an important mechanism for hadron
production in this environment. Recall that in p + p collisions, the proton to
pion ratio as a function of $p_t$ remains basically unchanged, never exceeding
one, for collision energies ranging from 19.4 GeV at the Tevatron, 44.6 and
52.8 GeV at ISR up to 200 GeV at RHIC~\cite{STAR}. This stable behavior
with collision energy can be understood as a manifestation of the fact that in
p + p collisions, hadron production is dominated by parton fragmentation and
the proton to pion ratio simply reflects the ratio of the 
corresponding fragmentation functions. In contrast, the proton to 
pion ratio in collisions of heavy systems grows with collision
energy~\cite{Laszlo,STAR2}, reaching and even exceeding one in Au + Au
collisions at $\sqrt{s_{\mbox{\tiny{NN}}}}=200$ GeV for $p_t\sim 2$
GeV/c. While it is possible to understand the different behaviors of 
the above ratios in collisions of heavy systems as partially arising from
the increase of radial flow with collision energy~\cite{ayala1, ayala2},
another important ingredient that needs to be understood is how these ratios
are influenced by the relative abundance of protons and pions when these
particles are produced in the collision of heavy systems at the highest
energies.

In its simplest form, the recombination scenario explains the formation of low
to intermediate $p_t$ hadrons from the bounding of quarks in a densely
populated phase space, assigning appropriate degeneracy factors for mesons and
baryons~\cite{recomb,Fries}. However, it only accounts for correlations among  
quarks in momentum space but misses describing the corresponding expected 
correlation in coordinate space. An implicit assumption is that hadronization
happens at a single temperature. However, it is known that hadronization is 
not an instantaneous process but rather that it spans a window of temperatures 
and densities. For instance lattice calculations~\cite{Karsch} show that the
phase transition from a deconfined state of quarks and gluons to a hadron gas
is, as a function of temperature, not sharp. The question that arises is to
what extent the probability to recombine quarks into mesons and baryons
depends on density and temperature and whether this probability differs for
hadrons with two and three constituents, that is to say, whether the relative
population of baryons and mesons can be attributed not only to the degeneracy
factors but rather to the dynamical properties of quark clustering in a
varying density scenario.

The detailed answer to the above question is a subject that belongs
to the realm of confinement phenomena and thus to the dynamics of
non-perturbative QCD. It is however possible to also address it in the context
of numerical models that reproduce aspects of QCD phenomenology such as quark
clustering at low density and color deconfinement at high density. One of such
models is the so called string-flip model~\cite{stringflip} which has proven
to be successful in the study of quark/hadron matter as a function of
density~\cite{string1,Genaro1,Genaro2}. Other approaches toward a dynamical 
description of recombination, in the context of fluctuations in heavy ion 
collisions, have been recently formulated in terms of the qMD 
model~\cite{0702188}.

In this work we compute the meson and baryon thermal spectra from a dynamical
model of quark recombination. The work is organized as follows: In
Sec.~\ref{II}, we introduce the formalism to incorporate the probability to
form clusters of two and three quarks as a function of density and
temperature in the description of thermal particle spectra. In Sec.~\ref{III}
we present the basics to implement the numerical simulations in the
string-flip model. In Sec.~\ref{IV} we use this model to compute the
probability to form mesons and baryons from liberated quarks. In Sec.~\ref{V}
we use these spectra to compute the proton to pion ratio and compare the
results to experimental data. We finally present our conclusions and give an
outlook in Sec.~\ref{concl}. 

\section{Thermal Spectra}\label{II}

In general, the invariant transverse momentum distribution of a given hadron
can be written as an integral over the freeze-out, space-time hypersurface
$\Sigma$ of the relativistically invariant phase space particle
density $F(x,P)$,
\be
   E\frac{dN}{d^3P}=g
   \int_{\Sigma_f}d\Sigma\ \frac{P\cdot u(x)}{(2\pi)^3}F(x,P)\, ,
   \label{distribution}
\ee
where $P$ is the hadron's momentum and $u(x)$ is a future oriented unit
four-vector normal to $\Sigma$ and $g$ is the degeneracy factor for the
hadron which takes care of the spin degree of freedom.

In the recombination model, the phase space particle density is taken as the
convolution of the product of Wigner functions for each hadron's constituent
quark at a given temperature and the constituent quark wave function inside
the hadron. For instance, the meson phase space distribution is given by
\be
   F^M(x,P)&=&\sum_{a,b}\int_0^1dz|\Psi_{ab}^M(z)|^2w_a({\mathbf{x}},zP^+)
   \nonumber\\
   &\times&\bar{w}_b({\mathbf{x}},(1-z)P^+)\, ,
   \label{wigmes}
\ee
where $P^+$ is the light-cone
momentum, $\Psi_{ab}^M(z)$ is the meson wave function and $a,\ b$ represent
the quantum  numbers (color, spin, flavor) of the constituent quark and
antiquark in the meson, respectively. An analogous equation can also be
written for baryons. When each constituent quark's Wigner function is
approximated as a Boltzmann distribution and momentum conservation is used,
the product of Wigner functions is given by a Boltzmann-like factor that
depends only on the light-cone momentum of the hadron~\cite{Fries}. For
instance, in the case of mesons 
\be
   w_a({\mathbf{x}},zP^+)\bar{w}_b({\mathbf{x}},(1-z)P^+)&\sim&
   e^{-zP^+/T}e^{-(1-z)P^+/T}\nonumber\\
   &=&e^{-P^+/T}\, .
\ee
In this approximation, the product of parton distributions is independent of
the parton momentum fraction and the integration of the wave function over $z$
is trivially found by normalization. There can be corrections 
from a dependence of each constituent quark Wigner function on momentum
components that are not additive because energy is not conserved in this
scenario~\cite{Fries2}. The QCD dynamics between quarks inside the hadron is
encoded in the wave function.  

In order to allow for a more realistic dynamical recombination scenario let us
take the above description as a guide, modifying the ingredients that account
for the QCD dynamics of parton recombination. Let us assume that the phase
space occupation can be factorized into the product of a term containing the
thermal occupation number, including the effects of a possible flow
velocity, and another term containing the system energy density $\epsilon$
driven probability ${\mathcal{P}}(\epsilon)$ of the coalescence of partons into
a given hadron. We thus write the analog of Eq.~(\ref{wigmes}) as  
\be
   F(x,P)=e^{-P\cdot v(x)/T}{\mathcal{P}}(\epsilon)\, ,
   \label{ourF}
\ee
where $v(x)$ is the flow velocity.

To compute the probability ${\mathcal{P}}(\epsilon)$ let us consider a model
that is able to provide information about the likelihood of clustering of
constituent quarks to form hadrons from an effective quark-quark
interaction. We will explicitly consider the so called string-flip model which
we proceed to describe. 

\section{String Flip Model}\label{III}

The String Flip Model is formulated incorporating a many-body quark potential
able to confine quarks within color-singlet clusters
\cite{stringflip}. At low densities, the model describes a given system of
quarks as isolated hadrons while at high densities, this system becomes a free
Fermi gas of quarks. For our purposes, we consider up and down
flavors and three colors (anticolors) quantum numbers. Our approach is very
close to that described in Refs.~\cite{string1} and~ \cite{Genaro1}, where
we refer the reader for an extensive discussion of the model details. For
completeness, here we describe the basic ingredients. 
 
\subsection{Many-body potential}\label{IIIa}

The many-body potential is defined as the optimal clustering of quarks into 
color-singlet objects, where by optimal we mean the configuration that
minimizes the potential energy. In our approach, the interaction between
quarks is pair-wise. Therefore, the optimal clustering is achieved by finding
the optimal pairing between two given sets of quarks of different color for all
possible color charges.
 
Consider, for example, a set of $A$ quarks some with color $c_1$ and others
with color $c_2$, irrespective of flavor, in accordance to the the
flavor-blindness nature of QCD. We define the optimal pairing between $c_1$
and $c_2$ quarks as:
\be
   V_{c_1 c_2}=\underset{P}{min}\sum\limits_{i=1}^{A}
   \textit{v}[r_{ic_1},P(r_{ic_2})],
   \label{Potencial}  
\ee
where $r_{ic_1}$ is the spatial coordinate of the $i$-th $c_1$ quark
and $P(r_{ic_2})$ is the coordinate of the mapped $j$-th $c_2$ quark. The
minimization procedure is performed over all possible  $N$! permutations of
the $c_1$ quarks and the interaction between quarks (\textit{v}) is assumed to
be harmonic:
\be
   \textit{v}(r_{ic_1},r_{jc_2})=\frac{1}{2}k(r_{ic_1}-r_{jc_2})^2\, ,
   \label{smalpot}
\ee 
where $k$ is the spring constant. Through this procedure, we can distinguish
two types of hadrons:  

i) {\it Meson-like}. In this case the pairing is imposed to be between color
and anticolors and the many-body potential of the system made up of mesons is
given by:
\be
   V_\pi = V_{B\bar{B}}+V_{G\bar{G}}+V_{R\bar{R}}\,
   \label{mespot}
\ee
where the individual terms are given by Eq.~\ref{Potencial} for the
corresponding colors. $R(\bar{R})$, $B(\bar{B})$ and $G(\bar{G})$ are the
labels for red, blue and green color (anticolor) respectively. Note that this
potential can only build pairs. 
 
ii){\it Baryon-like}. In this case the pairing is imposed to be between the
different colors in all the possible combinations. In this manner, the
many-body potential is:
\be
   V_p = V_{RB}+V_{BG}+V_{RG}\, 
   \label{barpot}
\ee
which can build colorless clusters by linking 3(RBG), 6(RBGRBG),... etc.,
quarks. Since the interaction is pair-wise, the 3-quark clusters are of the
delta (triangular) shape.
 
According to QCD phenomenology, the formed hadrons should interact
weakly due to the short-range nature of the hadron-hadron interaction. This is
partially accomplished by the possibility of a quark flipping from one cluster
to another. At high energy density, asymptotic freedom demands that quarks
must interact weakly. This behavior is obtained once the average inter-quark
separation is smaller than the typical confining scale. 
 
The Hamiltonian for the system built up of $N$ quarks of mass $m_i$ and
momentum ${\bf p}_i$ is given by:
\be
   H=\sum_{i=1}^{N}\frac{{\bf p}_i^2}{2m_i}+V(\textbf{x}_1,...,\textbf{x}_N).
   \label{Hamiltonian}
\ee
In this work, as a first approach, we will study the meson and baryon like
hadrons independently. Therefore, $V=V_\pi$ or $V_p$, depending on the type
of hadrons we wish to describe.
 
\subsection{The variational wave function}\label{IIIb}

We use a variational Monte Carlo approach to describe the evolution of a
system of $N$ quarks as a function of the particle density. We consider the
quarks  moving in a three-dimensional box whose sides has length \textit{a}
and the system described by a variational wave function of the form:
\be
   \Psi_{\lambda}(\textbf{x}_1,...,\textbf{x}_N)=e^{-\lambda
   V(\textbf{x}_1,...,\textbf{x}_N)}\Phi_{FG}(\textbf{x}_1,...,\textbf{x}_N),
   \label{wavefun}
\ee

where $\lambda$ is the single variational parameter,
$V$(\textbf{x}$_1$,...,\textbf{x}$_N$) is the many-body potential defined in 
Eqs.~(\ref{mespot}) and~(\ref{barpot}) for mesons and baryons respectively,
and $\Phi_{FG}$(\textbf{x}$_1$,...,\textbf{x}$_N$) is the Fermi-gas wave
function given by a product of Slater determinants, one for 
each color-flavor combination of quarks. These are built up from
single-particle wave functions describing a free particle in a box
\cite{Genaro1}. The selection of 
the variational wave function rests in the fact that we are interested in the
evolution of the system from low to high energy densities. The exponential
term is responsible of the clustering correlations. At low energy density, the
system is formed by isolated color-singlet hadrons and quarks strongly
interact inside each cluster; in this case, the exponential term of the wave
function has a big contribution since the average interquark distance is of
the order of the confining scale. In contrast, at high energy density, where
asymptotic freedom takes place, the interaction between quarks is weak and the
system looks like a Fermi gas of quarks. In this case, the inter-quark
separation is much smaller than the confining scale and the exponential term
effect vanishes. 

As we will show below, the variational parameter changes from a fixed value
at low energy density (isolated clusters) to zero at high energy density
(Fermi gas).
 
\subsection{Variational Monte Carlo Calculations}\label{IIIc}

We need to evaluate the expectation value of the Hamiltonian operator given by
Eq.~(\ref{Hamiltonian}) and minimize it with respect to the variational
parameter $\lambda$ 
\be
   \frac{\partial <\Psi_\lambda |H|\Psi_\lambda >}{\partial \lambda}&=0.
   \label{expect}
\ee
In order to do this, we can take advantage of the structure of the wave
function in Eq.~(\ref{wavefun}), which allows some simplifications in the 
calculations. After an integration by parts, the expectation
value of the Hamiltonian operator for a given particle density can be
evaluated as: 
\be
   <H>_{\lambda}=T_{FG}+2\lambda^2<W>_{\lambda}+<V>_{\lambda},
   \label{expecfordensity}
\ee
where $T_{FG}$ is the kinetic energy of a free Fermi gas,
$<V>_{\lambda}$ is the potential energy as defined by Eqs.~(\ref{mespot})
and~(\ref{barpot}) and $<W>_{\lambda}$ is the term that indicates how the
kinetic energy increases due to clustering correlations. This last is
expressed as: 
\be
   <W>_\lambda=\sum_i^N\frac{1}{m_i}(\textbf{x}_i-\textbf{y}_i)^2,
   \label{kinterm}
\ee
where the sum is over all quarks in the system and \textbf{y}$_i$ represents
the average position of the two quarks connected to the $i$-th quark for a
baryon-like hadron or the corresponding antiquark position for a meson-like
hadron.
 
A very important property of having written the expectation value
$<H>_{\lambda}$ as in Eqs.~(\ref{expecfordensity}) and~(\ref{kinterm}) is that
the two functions to be evaluated ($V$ and $W$) are local; consequently, their
expectation values can be calculated 
using Monte Carlo techniques, in particular we use the Metropolis method to
do the sampling. Finally, this procedure is carried out for different values
of the particle density.

\begin{figure}[t!] 
{\centering
{\epsfig{file=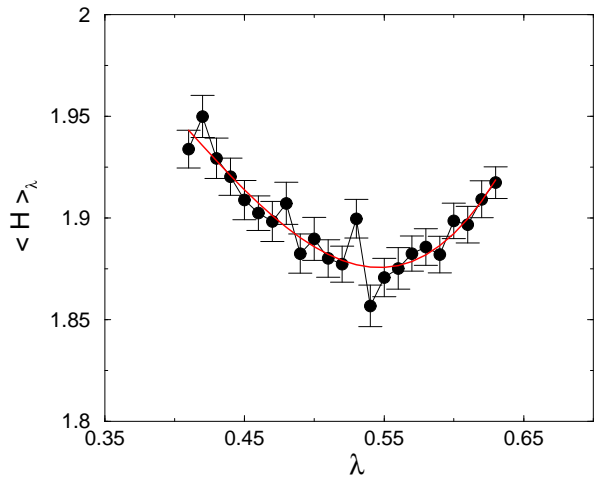, width=0.9\columnwidth}}
\par}
\caption{(Color on line) Energy per particle from the simulation (symbols) as
  a function of 
  the variational parameter at  low density $\rho = 0.17\rho_0$ for the baryon
  system. The solid line is a fit to the data.}
\label{energyzero}
\end{figure}

The variational parameter has definite values for the extreme density cases.
At very low density it must correspond to the wave function solution of an
isolated hadron. For example, the non-relativistic quark model
for a hadron consisting of 2 and 3 quarks, bound by a harmonic potential,
predicts, in units where $k=m=1$ that $\lambda_\pi \to \lambda_{0\pi} =
\sqrt{1/2}$ and $\lambda_p \to \lambda_{0p} = \sqrt{1/3}$ respectively; at very
high densities the value of $\lambda$ must vanish for both cases. In the
appendix, we show how to convert from the system of units where $k=m=1$ to
physical units.

\section{Probabilities}\label{IV}

All the results we present here come from simulation done with 384 particles,
192 quarks and 192 antiquarks, corresponding to having 32 $u \ (\bar{u})$ plus
32 $d\ (\bar{d})$ quarks  (antiquarks) in the three color charges
(anti-charges).  

To determine the variational parameter as a function of density we proceed as
follows: first we select the value of the particle density $\rho$ in the box,
which, for a fixed number of particles, means changing the box size. Then we
compute the energy of the system as a function of the variational parameter
using the Monte Carlo Method described in the previous section. The minimum of
the energy determines the optimal variational parameter. We repeat the
procedure for a 
set of values of the particle densities in the region of interest. In fig.
\ref{energyzero} we exemplify the behavior of the energy per particle from the
simulation (symbols) as a function of the variational parameter at low density
($0.17 \rho_0$) for a baryon system. The solid line is a fit to the data. We
can observe that the expected values from the solution to the 
3-quark system with a harmonic potential are properly approached. This is
taken as an indication of the range where the value of the variational
parameter should lay for an isolated hadron.

\begin{figure}[t!] 
{\centering
{\epsfig{file=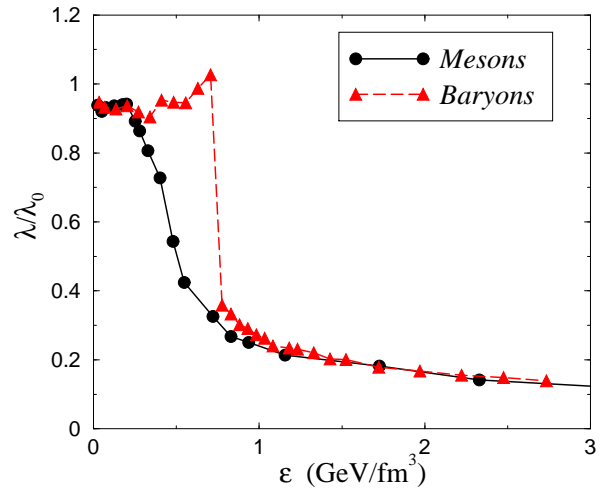, width=0.9\columnwidth}}
\par}
\caption{(Color on line) Normalized variational parameter as a function of the
  energy density for the baryon (triangle symbols) and meson (circle symbols)
  cases.}
\label{lambda2}
\end{figure}

In order to check whether the variational approach describes the
expected results in the low and high density limits,  we display
in fig.~\ref{lambda2} the energy density dependence of the variational
parameter $\lambda$  for the baryon, $\lambda_p$, (triangle symbols) and
meson, $\lambda_\pi$, (circle symbols) cases. In both cases, the 
behavior of the optimal parameter is the expected one, as described in detail
in the previous section.  At low densities, $\lambda$
converges to the theoretical value ($\lambda_{0\pi}= 2^{-1/2}$ and
$\lambda_{0p}= 3^{-1/2}$) for baryons and mesons respectively. We
use these to normalize the data such that the values for low energy densities
is around 1; at high densities the values of both $\lambda_p$ and
$\lambda_\pi$ approaches zero 
asymptotically. Nevertheless, there are differences between the two cases. In
the case of baryons, there is a deep drop around 0.7 GeV/fm$^3$ indicating
that the length scale for quark confinement has increased in the medium (this
unexpected behavior was also reported in Refs.~\cite{Genaro1,string1}), to
subsequently evolve into the Fermi gas domain. For a meson system, the
variational parameter has a smooth drop. 
The above indicates a qualitative difference in the formation of
baryon and meson clusters, as a function of the energy density.

\begin{figure}[t!] 
{\centering
{\epsfig{file=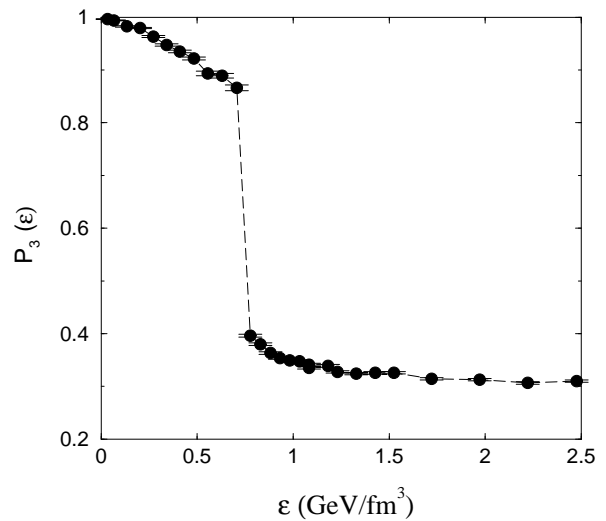, width=0.9\columnwidth}}
\par}
\caption{(Color on line) Percentage of clusters of 3 quarks  for different
  energy densities.}  
\label{P3}
\end{figure}

The information contained in the variational parameter is global, in
the sense that it only gives an approximate idea about the average size of the
inter-particle distance at a given density, which is not necessarily the same
for quarks in a single cluster. This is reflected in the behavior of the
variational parameter $\lambda_p$ for 
the case of baryons which, from fig.~\ref{lambda2}, goes above 1 for energies
close to where the sudden drop in the parameter happens. We interpret this
behavior as as a consequence of the procedure we employ to produce colorless
clusters for baryons, which, as opposed to the case to form mesons, allows the
formation of clusters with a number of quarks greater than 3. When including
these latter clusters, the information on their size is also contained in
$\lambda$. To correct for this, we compute the likelihood to find
clusters of 3 quarks $P_3$. Recall that for $3N$ quarks in the system, the
total number of clusters of 3 quarks that can be made is equal to $N$. However
this is not always the case as the density changes, given that the potential
allows the formation of clusters with a higher number of quarks. $P_3$ is 
defined as the ratio between the number of clusters of 3 quarks found at a
given density, with respect to $N$. This is displayed in fig.~\ref{P3}. 

\begin{figure}[t!] 
{\centering
{\epsfig{file=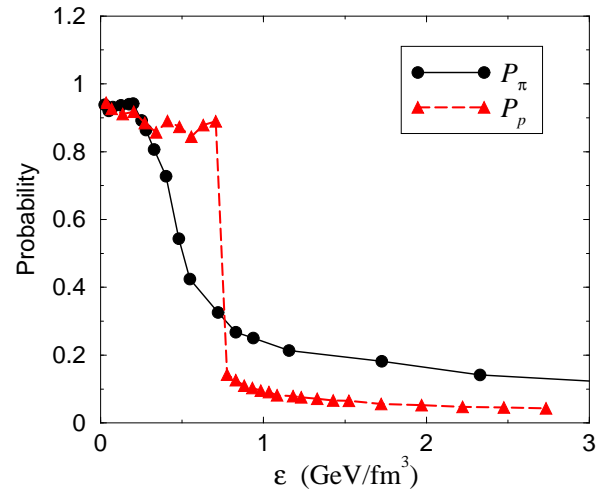, width=0.9\columnwidth}}
\par}
\caption{(Color on line) Probabilities to form baryons and mesons as a
  function of energy density.}   
\label{P_pP_pi}
\end{figure}

Therefore, within our approach, we can define
the probability of forming a baryon as the product of the
$\lambda/\lambda_{0p}$ parameter times $P_3$, namely
\be
   {\mathcal P}_p=\lambda/\lambda_{0p} \times P_3.
   \label{probprot} 
\ee
For the case of mesons, since the procedure only takes into account the
formation of colorless quark-antiquark pairs, we simply define the probability
of forming a meson as the value of the corresponding normalized variational
parameter, namely
\be
   {\mathcal P}_\pi=\lambda/\lambda_{0\pi}.
   \label{probmes}
\ee
The probabilities ${\mathcal P}_p$ and ${\mathcal P}_\pi$ as a function of the
energy density are displayed in fig.~\ref{P_pP_pi}. Notice the qualitative
differences between these probabilities. In the case of baryons, the sudden
drop found in the behavior of the variational parameter is preserved at an
energy density around $\epsilon =0.7$ GeV/fm$^3$ whereas in the case of
mesons, this probability is smooth, indicating a difference in the production
of baryons and mesons with energy density.

These are the probabilities we set originally ourselves to look for and that
we use in sec~\ref{V} to compute the proton to pion ratio. Before proceeding
to that analysis, it is also instructive to learn more about the behavior of
other properties of the quark system with energy density.  

\subsection{Clusters' size}

To obtain a more precise information about the clusters' size, we can look at
the distribution of particles into a cluster, characterized by a mean square
radius (MSR). To do so, we proceed to calculate the frequency distribution of
clusters with a given value of the MSR. 
For the baryon system, this corresponds to first identify the clusters made of
3 quarks --given that clusters with more than 3 quarks are allowed to form--
and then to compute their MSR. For the meson system, as
we allowed only the formation of clusters with 2 quarks, we compute the MSR
for all the quark pairs as selected by the optimal pairing search. 

\begin{figure}[t!] 
{\centering
{\epsfig{file=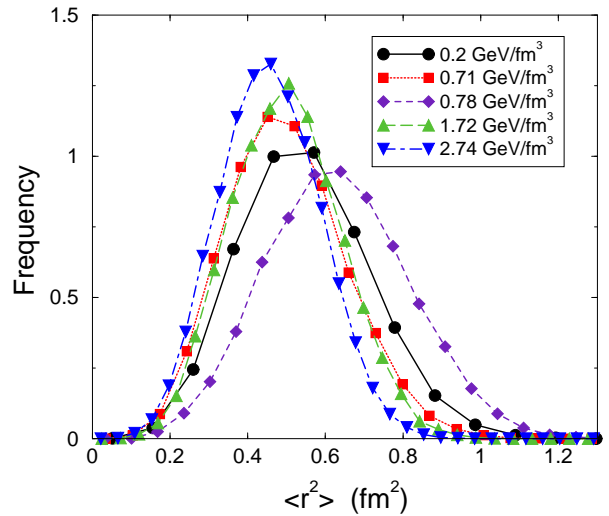, width=0.9\columnwidth}}
\par}
\caption{(Color on line) Normalized MSR frequency distributions for baryons
  for a set of energy densities.}   
\label{msrB}
\end{figure}

\begin{figure}[t] 
{\centering
{\epsfig{file=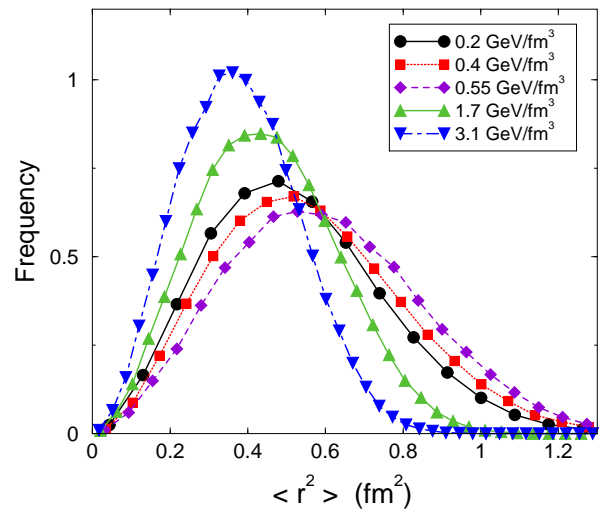, width=0.9\columnwidth}}
\par}
\caption{(Color on line) Normalized MSR frequency distributions for mesons for
  a set of energy densities.}
\label{msrM}
\end{figure}

The MSR distributions for a set of energy densities are shown in
figs.~\ref{msrB} and~\ref{msrM} for the baryon and meson systems,
respectively. They have been normalized such that their corresponding area
under the curve is one. 
It can be noticed, in both cases, that  the distribution becomes narrower as
the density grows. The physical meaning is that as the system gets denser,
quarks are brought closer to each other. This is precisely the expected
behavior when the QGP is forming.

\begin{figure}[t!] 
{\centering
{\epsfig{file=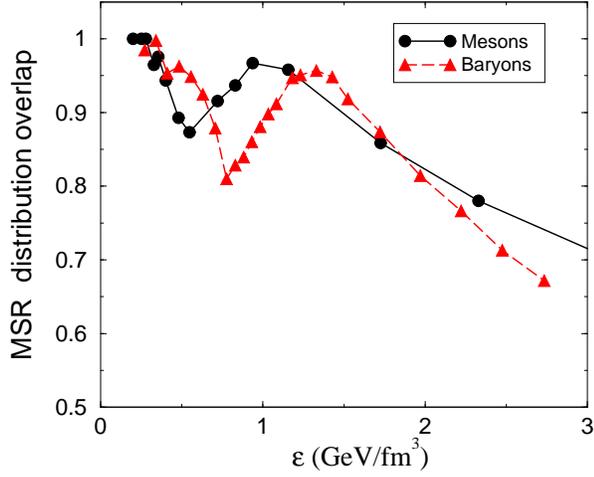, width=0.9\columnwidth}}
\par}
\caption{(Color on line) Overlap function describing how much the clusters
  (mesons and baryons), resemble the MSR distribution of the one computed at
  normal nuclear energy density.}  
\label{Pr}
\end{figure}

The overlap between the distributions of particle clusters MSR at high
densities and the distribution at low density (we take $\rho_0=0.2$ GeV/fm$^3$
as reference), indicates how much the distribution at a given density looks
like the distribution at the lowest density. This overlap, as a function of
density, is shown in fig.~\ref{Pr} for 
both  baryons and mesons. Notice that in the interval 0.35 $\leq\rho\leq$ 1
GeV/fm$^3$, this overlap in the case of baryons is larger than the overlap in 
the case of mesons while in the interval 1 $\leq\rho\leq$ 1.4 GeV/$fm^3$ the
behavior is the opposite. We interpret this as a further indication of a
difference in the production of baryons and mesons as the phase transition is
taking place. 

\subsection{Correlation functions}

\begin{figure}[t!] 
{\centering
{\epsfig{file=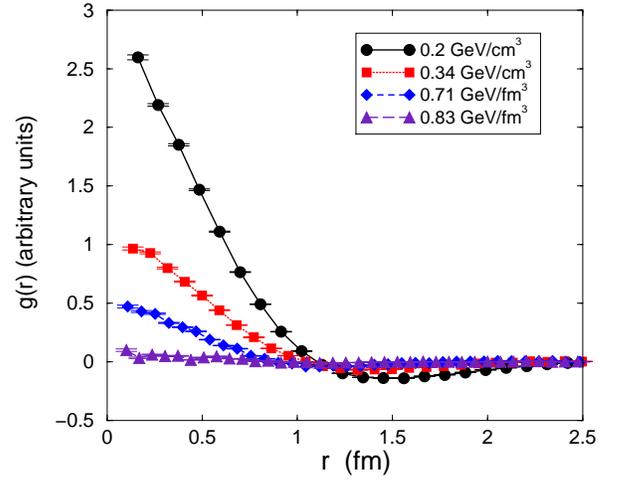, width=0.9\columnwidth}}
\par}
\caption{(Color on line) Baryon correlation function for a set of energy
  densities.}   
\label{gB}
\end{figure}

\begin{figure}[t!] 
{\centering
{\epsfig{file=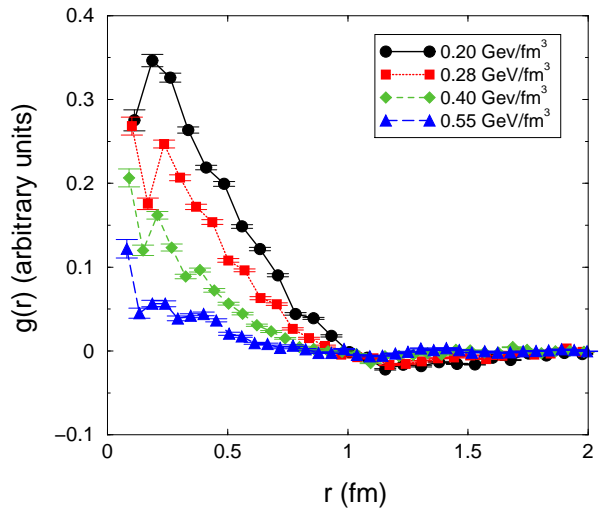, width=0.9\columnwidth}}
\par}
\caption{(Color on line) Meson correlation function for a set of energy
  densities.}
\label{gM}
\end{figure}

In order to further explore the meaning of this result, we have computed the
two-body correlation function, which measures the probability of finding  two
given quarks at a relative distance \textbf{r}. This function is defined as
\cite{walecka} 
\be
   \rho{\hbox{\lower 3pt \hbox{${_2}$}}}({\bf r}) =
   \sum_{\alpha\beta} 
   \langle\Psi_{0}|
   \hat{\psi}_{\alpha}^{\dagger}({\bf r})
   \hat{\psi}_{\beta}^{\dagger}({\bf 0})
   \hat{\psi}_{\beta}({\bf 0})
   \hat{\psi}_{\alpha}({\bf r})|\Psi_{0}\rangle \;,
 \label{rho2}
\ee
where $\alpha,\beta$ denote the collection of all internal quantum numbers,
such as color and flavor. The importance of this observable is that its
behavior can tell us if there is any correlation between quarks without
relaying to the pairing information. For example, the two-body correlation
function between quarks of different colors at low density, where quarks are in
isolated hadrons, will show an increase for distances around the hadron size
while for high densities it won't be able to identify a particular distribution
and therefore it will become flat. It is worth to mention that if we compute
the correlation between quarks of the same color, the shape will reflect the
statistical correlation between fermions by a dip at short
distances~\cite{Genaro1}. 
Figures~\ref{gB} and~\ref{gM} show the correlation function
between quarks of different color for a set of energy densities, for a baryon
and meson systems, respectively. It is common to normalize the correlation
function such that the long distance behavior converges to one. However, we
have chosen the normalization such that the uncorrelated long distance
behavior converges to zero.

\begin{figure}[t!] 
{\centering
{\epsfig{file=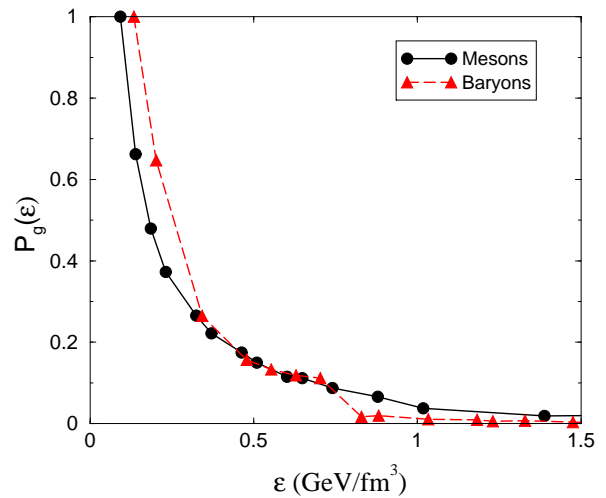, width=0.9\columnwidth}}
\par}
\caption{(Color on line) Normalized correlation function for mesons and
  baryons as a function of the energy density. See text for details.}
\label{gMB}
\end{figure}

To quantify the changes in the shape of the correlation function we define the
function $P_g$, as 
\be
   P_g=\frac{\int g(r,\rho) dr}{\int g(r,\rho_0)dr }
   \label{corrprob}
\ee 
where we have again chosen the normal nuclear matter density $\rho_0$ as the
reference. We display in fig.~\ref{gMB} the behavior of such function for
mesons (circle symbols) and baryons (triangle symbols). Both systems exhibit a
rapid evolution to low values well before the energy density reaches the
region where the variational parameter drops down, although there is a
small change for the baryon system in that energy (around 0.7 \ GeV/fm$^3$),
and then they evolve to become practically negligible around 1.5 \ GeV/fm$^3$. 

\section{proton to pion ratio}\label{V}

In order to quantify how the different probabilities to produce sets of three
quarks (protons) as compared to sets of two quarks (pions) affect these
particle's yields as the energy density changes during hadronization, we need
to resort to a model for the space-time evolution of the collision. For the present purposes, we will omit describing the effect of radial flow and take
Bjorken's scenario which incorporates the fact that initially, expansion is
longitudinal, that is, along the beam direction which we take as the $\hat{z}$
axis. In this 1+1 expansion scenario, the relation between the temperature $T$
and the 1+1 proper-time $\tau$ is given by 
\be
   T=T_0\left(\frac{\tau_0}{\tau}\right)^{v_s^2},
   \label{temperaturevstau}
\ee
where $\tau=\sqrt{t^2-z^2}$. Equation~(\ref{temperaturevstau}) assumes that
the speed of sound $v_s$ changes slowly with temperature. A lattice estimate
of the speed of sound in quenched QCD~\cite{Gupta} shows that $v_s^2$
increases monotonically from about half the ideal gas limit 
for $T\gtrsim 1.5 T_c$ and approaches this limit only for
$T>4T_c$, where $T_c$ is the critical temperature for the phase transition. No
reliable lattice results exist for the value of the speed of sound in the
hadronic phase though general arguments indicate that the equation of state
might become stiffer below $T_c$ and eventually softens as the temperature
approaches zero~\cite{McLerran}. For the ease of the argument, here we
take $v_s$ as a constant equal to the ideal gas limit $v_s^2=1/3$.  

We also consider that hadronization takes place on hypersurfaces $\Sigma$
characterized by a constant value of $\tau$ and therefore
\be
   d\Sigma=\tau\rho \ d\rho \ d\phi\  d\eta ,
   \label{hypersurface}
\ee
where
\be
   \eta=\frac{1}{2}\ln\frac{t+z}{t-z},
   \label{spacerapidity}
\ee
is the spatial rapidity and $\rho$, $\phi$ are the polar transverse
coordinates. Thus, the transverse spectrum for a hadron species $H$
is given as the average over the hadronization interval of the right hand-side
of Eq.~(\ref{distribution}), namely
\be
   E\frac{dN^H}{d^3P}=\frac{g}{\Delta \tau}
   \int_{\tau_0}^{\tau_f}d \tau\int_{\Sigma}d\Sigma\ \frac{P\cdot
     u(x)}{(2\pi)^3}F^H(x,P), 
   \label{distributionaveraged}
\ee
where $\Delta \tau=\tau_f-\tau_0$.

To find the relation between the energy density $\epsilon$ --that the
probability ${\mathcal{P}}$ depends upon-- and $T$, we resort to lattice
simulations. For the case of two flavors, a fair representation
of the data~\cite{Karsch} is given by the analytic expression
\be
   \epsilon /T^4 = a\left[ 1 + \tanh\left(\frac{T-T_c}{bT_c}\right)\right],
   \label{latticeenergy}
\ee
with $a=4.82$ and $b=0.132$. We take $T_c=175$ MeV.

For a purely longitudinal expansion, the flow four-velocity vector $v^\mu$ and
the normal to the freeze-out hypersurfaces of constant $\tau$, $u^\mu$,
coincide and are given by
\be
   v^\mu=u^\mu=(\cosh\eta,0,0,\sinh\eta),
   \label{flowvel}
\ee
therefore, the products $P\cdot u$ and $P\cdot v$ appearing in
Eq.~(\ref{distributionaveraged}) can be written as
\be
   P\cdot v=P\cdot u=m_t\cosh(\eta-y),
   \label{Pdotv}
\ee
where $m_t=\sqrt{m_H^2+p_t^2}$ is the transverse mass of the hadron and
\be
   y=\frac{1}{2}\ln\left(\frac{E+p_z}{E-p_z}\right)
   \label{rapidity}
\ee
is the rapidity.

Considering the situation of central collisions, we can assume that there is
no dependence of the particle yield on the transverse polar
coordinates. Integration over these variables gives thus the transverse
overlap area of the colliding nuclei $A$, namely
\be
   \int\rho d\rho d\phi \rightarrow A.
   \label{intoverrhophi}
\ee

Looking only at the case of central rapidity, $y=0$, integration over
the spatial rapidity variable can also be performed straightforward. The
integral involved is
\be
   \int_{-\infty}^{\infty}d\eta e^{-(P\cdot v)/T}P\cdot u=
   2m_tK_1(x),
   \label{intovereta}
\ee
where $x\equiv m_t/T$ and $K_1$ is a Bessel function of the second kind
and first order. Gathering the above elements, the
final expression for the hadron's transverse distribution is given by
\be
   E\frac{dN^H}{d^3P}=\frac{g}{(2\pi)^3}\frac{2m_tA}{\Delta \tau }
   \int_{\tau_0}^{\tau_f}d \tau\tau K_1\left[\frac{m_t}{T(\tau )}\right]
   {\mathcal{P}}[\epsilon (\tau )].
   \label{distfin}
\ee

\begin{figure}[b!] 
{\centering
{\epsfig{file=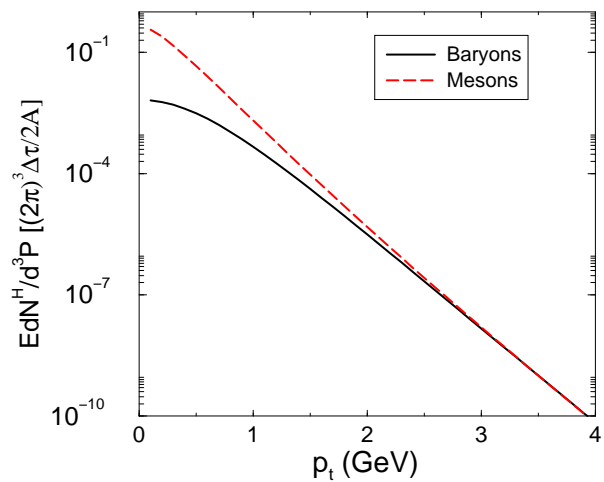, width=0.9\columnwidth}}
\par}
\caption{(Color on line) Proton and pion invariant distributions as a function
  of transverse momentum for $\tau_0=0.75$ fm, $\tau_f=3.5$ fm and $T_0=200$
  MeV, corresponding to a final freeze-out temperature of $\sim 120$ MeV.}
\label{fig11}
\end{figure}

\begin{figure}[t!] 
{\centering
{\epsfig{file=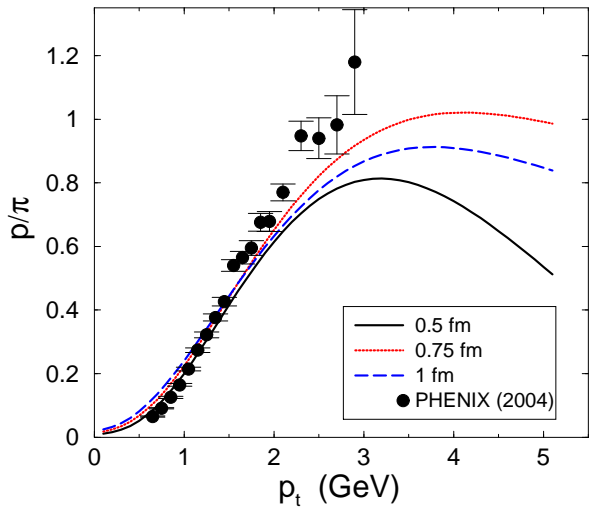, width=0.9\columnwidth}}
\par}
\caption{(Color on line) Proton to pion ratio as a function of transverse
  momentum for three different values of the initial evolution proper-time
  $\tau_0=0.5,\ 0.75$ and $1$ fm and the same finial freeze-out proper-time
  $\tau_f=3.5$ fm, compared to data for this ratio for Au + Au collisions at
  $\sqrt{s_{NN}}=200$ GeV from PHENIX. The height of this ratio is very
  sensitive to the choice of the initial evolution time.} 
\label{fig12}
\end{figure}

Figure~\ref{fig11} shows an example of the the pion and proton distributions
obtained by means of Eq.~(\ref{distfin}), where we have used the values
$\tau_0=0.75$ fm and $\tau_f=3.5$ fm and an initial temperature $T_0=200$
MeV. From Eq.~(\ref{temperaturevstau}), this corresponds to a final 
freeze-out temperature of $\sim 120$ MeV. For protons we take a degeneracy
factor $g=2$ whereas for pions $g=1$, to account for the spin degrees of
freedom. Figure~\ref{fig12} shows the proton 
to pion ratio for three different values of the initial evolution proper time
$\tau_0=0.5,\ 0.75$ and $1$ fm and the same finial freeze-out proper-time
$\tau_f=3.5$ fm, compared to data for this ratio for Au + Au collisions at
$\sqrt{s_{NN}}=200$ GeV from PHENIX~\cite{PHENIXBM}. We notice that the
maximum height reached by this ratio is sensitive to the choice of the initial
evolution time. We also notice that the $p_t$ value for which the maximum is
reached is displaced to larger values than what the experimental values
indicate. This result is to be expected since the model assumptions leading to
Eq.~(\ref{distfin}) do not include the effects of radial flow that, for a
common flow velocity, are known to be larger for protons than for pions, and
which will produce the displacement of the ratio toward lower $p_t$ values. 

\section{Summary and Conclusions}\label{concl}

In conclusion, we have used the string-flip model to introduce a dynamical
quark recombination scenario that accounts for the evolution of the
probability to form a meson or a baryon as a function of the energy
density during the collision of a heavy-ion system. We have used the model
variational parameter as a measure of the probability to form colorless
clusters of three quarks (baryons) or of quark-antiquark (mesons). We have
shown that these probabilities differ; whereas the probability to form a pion
transits smoothly from the high to the low energy density domains, the
probability to form a baryon changes abruptly at a given critical energy
density. Within our approach, we attribute this difference to the way the
energy is distributed during the formation of clusters: whereas for mesons the
clustering happens only for quark-antiquark pairs, for baryons the energy can
be minimized by also forming sets of three, six, etc., quarks in (colorless)
clusters. These produces competing minima in the energy that do not reach each
other smoothly. We interpret this behavior as a signal for a qualitative
difference in the probability to form mesons and a baryons during 
the collision evolution. 

Notice that this approach is quantitatively different to the conventional coalescence scenario~\cite{recomb} where the probability to form a bound state of quarks is simply given by the overlap of the quark distribution functions with the Wigner function of the formed hadron and thus ignores the underlying evolving dynamics of the collision process. 

We have incorporated these different probabilities to
compute the proton and pion spectra in a thermal model for a Bjorken-like
scenario. We use these spectra to compute the proton to pion ratio as a
function of transverse momentum and compare to experimental data at the
highest RHIC energies. We argue that the ratio computed from the model is able
to reach a height similar to the one shown by data, although the maximum is
displaced to larger $p_t$ values. This could be understood by recalling that
the model does not include the effects of radial flow which is known to be
stronger for protons (higher mass particles) than pions. The inclusion of
these effects is the subject of current research that will be reported
elsewhere.

\section*{Appendix: Physical units}

The simulation was performed taking $m=k=1$. Here we show the
conversion to physical units.
 
Baryons:
To fix the the energy unit we first notice that in a 3-body system the energy
per particle, including its mass, is given by (with $m=k=1$): 
\be
\frac{E}{3}= \sqrt{3}+1.
\label{A1}
\ee
If we identify the state as the proton of mass $M_p=938$ MeV, then the
correspondence is 
\be
\sqrt{3}+1 \rightarrow 312.7\ {\mbox {MeV}}.
\label{A2}
\ee 
To fix the length unit we use the mean square radius, which for a 3-body
system is: $\sqrt{<r^2>}=(3)^{1/4}$. The experimental value for the proton
is
\be
\sqrt{<r^2>}=0.880 \pm 0.015\ {\mbox {fm}}.
\label{A3}
\ee 
Then the correspondence is: $(3)^{1/4} \rightarrow 0.88$ fm.

Mesons:
In a similar fashion we obtain for mesons (taking the pion as the
representative 2-body particle): 
Energy: $\frac{3}{2\sqrt{2}}+1 \rightarrow 70$ MeV, 
length: $2^{1/4} \rightarrow  0.764$ fm.

\section*{Acknowledgments}

Support for this work has been received by PAPIIT-UNAM
under grant number IN107105 and CONACyT under grant number
40025-F. M. Martinez was supported by DGEP-UNAM.

\end{document}